\documentclass[aps,prl,showpacs,superscriptaddress,reprint,amsmath,amssymb]{revtex4-1}
\usepackage{graphicx,color}

\begin{document}

\title{Increasing stripe-type fluctuations in $A$Fe$_{2}$As$_{2}$ ($A$ = K, Rb, Cs) superconductors probed by $^{75}$As NMR spectroscopy}

\author{Z. T. Zhang}
\affiliation{Hochfeld-Magnetlabor Dresden (HLD-EMFL), Helmholtz-Zentrum Dresden-Rossendorf, D-01314 Dresden, Germany}
\affiliation{Key Laboratory of Materials Physics, Institute of Solid State Physics, Chinese Academy of Sciences, Hefei 230031, People's Republic of China}

\author{D. Dmytriieva}
\author{S. Molatta}
\author{J. Wosnitza}
\affiliation{Hochfeld-Magnetlabor Dresden (HLD-EMFL), Helmholtz-Zentrum Dresden-Rossendorf, D-01314 Dresden, Germany}
\affiliation{Institut f\"{u}r Festk\"{o}rperphysik, TU Dresden, D-01060 Dresden Germany}

\author{S. Khim}
\author{S. Gass}
\author{A. U. B. Wolter}
\author{S. Wurmehl}
\author{H.-J. Grafe}
\email[Corresponding author. E-mail: ]{h.grafe@ifw-dresden.de}
\affiliation{IFW Dresden, Institute for Solid State Research, D-01171 Dresden, Germany}

\author{H. K\"{u}hne}\email[Corresponding author. E-mail: ]{h.kuehne@hzdr.de}
\affiliation{Hochfeld-Magnetlabor Dresden (HLD-EMFL), Helmholtz-Zentrum Dresden-Rossendorf, D-01314 Dresden, Germany}

\date{\today}

\begin{abstract}
We report $^{75}$As nuclear magnetic resonance measurements on single crystals of RbFe$_{2}$As$_{2}$ and CsFe$_{2}$As$_{2}$. Taking previously reported results for KFe$_{2}$As$_{2}$ into account, we find that the anisotropic electronic correlations evolve towards a magnetic instability in the $A$Fe$_{2}$As$_{2}$ series (with $A$ = K, Rb, Cs). Upon isovalent substitution with larger alkali ions, a drastic enhancement of the anisotropic nuclear spin-lattice relaxation rate and decreasing Knight shift reveal the formation of pronounced spin fluctuations with stripe-type modulation. Furthermore, a decreasing power-law exponent of the nuclear spin-lattice relaxation rate $(1/T_{1})_{H\parallel{ab}}$, probing the in-plane spin fluctuations, evidences an emergent deviation from Fermi-liquid behavior. All these findings clearly indicate that the expansion of the lattice in the $A$Fe$_{2}$As$_{2}$ series tunes the electronic correlations towards a quantum critical point at the transition to a yet unobserved, ordered phase.

\end{abstract}
\maketitle

The investigation of quantum critical points (QCPs) in iron pnictides provides a route for understanding the relation of their relatively high superconducting transition temperatures and the anomalous normal-state properties, such as enhanced quasiparticle masses and non-Fermi-liquid behavior \cite{BaFeAsP-QCP-London,Shibauchi-QCP-review,BaFeAsP-QCP-dHvA,BaFeAsP-QCP-nFL-RT, BaFeAsP-QCP-NMR,BaFeAsP-FeCo-QCP-NMR,QCP-JPCM-QSi,QCP-Eilers,QCP-Mizukami,Wang-K122-pressure,QCP-theory,QCP-theory}. 
Signatures of quantum criticality, driven by variation of the charge carrier concentration, have been reported for moderately Ni and Co doped BaFe$_{2}$As$_{2}$ \cite{Zhou2013, Ning2010, Chu2009}.
Also, by variation of the internal pressure, a QCP, emerging from the interplay of electronic localization and itinerancy, was theoretically anticipated \cite{QCP-theory}, and has been experimentally identified in BaFe$_{2}$(As$_{1-x}$P$_{x}$)$_{2}$ by various techniques \cite{BaFeAsP-QCP-London,Shibauchi-QCP-review,BaFeAsP-QCP-dHvA,BaFeAsP-QCP-nFL-RT,BaFeAsP-QCP-NMR,BaFeAsP-QCP-NMR}. 

In the case of the heavily hole-doped $A$Fe$_{2}$As$_{2}$ ($A$ = K, Rb, Cs) series, the Fe $3d$ orbitals are nominally filled with $N=5.5$ electrons.  
The proximity to the 
Mott-insulating phase, expected for $N=5.0$, promotes phenomena that arise from localization and increased electronic correlations.
Specifically, it was proposed that a QCP is approached with increasing chemical pressure, driven by an isovalent substitution of K$^{+}$ with Rb$^{+}$ or Cs$^{+}$ \cite{QCP-Eilers,QCP-Mizukami,Wang-K122-pressure}. In KFe$_{2}$As$_{2}$, a pronounced mass enhancement due to strong correlation effects, as well as deviations from standard Fermi-liquid behavior were reported from several studies \cite{Terashima2010, K122-crossover,Werner-NPhys-crossover,Liu-Lifshitz-crossover, Werner-NPhys-crossover,WQYu-K122,Dong-Non-fermi,Dong-Non-fermi}. An increase of the negative pressure in RbFe$_{2}$As$_{2}$ and CsFe$_{2}$As$_{2}$ does not affect the Fermi surface topology, but leads to enhanced quasiparticle masses, presumably driven by an orbital-selective increase of correlations \cite{Hardy2016,K122-crossover,Zhang-RbNonfermi,Rb122-gamma180,QCP-Eilers,Backes-KRbCs-Hund}. In line with these findings, the application of positive pressures was reported to decrease the Sommerfeld coefficient $\gamma$ in KFe$_{2}$As$_{2}$ \cite{K122-press-gamma}.
In a recent dilatometry and local-density approximation (LDA) study of the $A$Fe$_{2}$As$_{2}$ series, Eilers et al. suggested that the enhancement of correlations originates from a pressure-dependent hybridization of the in-plane Fe 3$d_{xy}$ orbitals \cite{QCP-Eilers}. The low-energy fluctuations close to the QCP seem to suppress $T_c$, in contrast to the usually propagated picture of superconductivity stimulated by critical fluctuations. 

As for the yet unobserved, ordered phase beyond the QCP, a combination of components from spin, orbital and charge degrees of freedom was proposed in a recent work by Drechsler et al. \cite{Drechsler2017}. For a $d_{x^2-y^2}$ superconducting order parameter, related critical spin fluctuations arising from the Fe 3$d_{xz}$/$d_{yz}$ bands could eventually enhance the quasiparticle masses without strengthening the formation of superconductivity. In support of the multi-component order, a tendency for charge order was experimentally observed in nuclear magnetic resonance (NMR) and nuclear quadrupole resonance (NQR) studies of KFe$_{2}$As$_{2}$ and RbFe$_{2}$As$_{2}$ \cite{Wang-K122-pressure, Civardi2016}.  


In general, approaching a magnetic QCP is reflected by non-Fermi-liquid behavior and a drastic increase of spin fluctuations, yielding a power-law behavior of related physical quantities with exponents differing from those of a Fermi liquid \cite{RMP-QCP-NonFermi}. Whereas the findings of Eilers et al. clearly indicate that a quantum critical point is approached with increasing FeAs-cell volume, other experimental reports seem much less conclusive. Different power-law dependencies between $\propto T^{1.5}$ and $\propto T^{2.0}$ were reported for the resistivity of 
the $A$Fe$_{2}$As$_{2}$ series \cite{K122-crossover,Dong-Non-fermi,Zhang-RbNonfermi,Hong-CsNonfermi,Khim2016}. 
From NMR measurements of all three $A$Fe$_{2}$As$_{2}$ compounds with  $H\parallel{c}$,
the same power law of $1/T_{1} \propto T^{0.75}$
was reported, i.e., giving no indication of approaching quantum criticality \cite{Wu-XHChen}.
In order to resolve this discrepancy between the different experimental findings, we performed a detailed NMR study of the anisotropic magnetic correlations.

In this paper, we present $^{75}$As NMR measurements on high-quality single crystals of RbFe$_{2}$As$_{2}$ and CsFe$_{2}$As$_{2}$ for both in-plane ($H\parallel{ab}$) and out-of-plane ($H\parallel{c}$) orientations of the magnetic field and temperatures between 1.6 and 300 K. Taking previously reported results for KFe$_{2}$As$_{2}$ into account, we find a continuous reduction of the Knight shift and a drastic increase of spin fluctuations along the $A$Fe$_{2}$As$_{2}$ series. Based on the anisotropy of $1/T_{1}$, we provide evidence 
for a stripe-type modulation of the spin fluctuations in all three $A$Fe$_{2}$As$_{2}$ compounds. Furthermore, a successively reduced power-law exponent of  $(1/T_{1})_{H\parallel{ab}}$ $\propto T^{\eta}$ at low temperatures shows increasing deviation from the unity exponent of a Fermi liquid. Our results clearly indicate that the electronic correlations are evolving from proximity to a Fermi liquid towards a state with strong correlations in close vicinity to a QCP.

Single crystals of $A$Fe$_{2}$As$_{2}$ ($A$ = Rb, Cs) were grown by use of the self-flux technique \cite{Khim2016}. Their stoichiometry was confirmed by energy-dispersive x-ray spectroscopy. For a precise sample orientation with field, a single-axis goniometer with a resolution of $0.1^\circ$ was used. The $^{75}$As NMR central transition spectra and nuclear spin-lattice relaxation rate $1/T_{1}$ were measured at temperatures from 1.6 to 300 K at $\mu_0 H=5.513$ T for RbFe$_{2}$As$_{2}$ and at $\mu_0 H=5.900$ T for CsFe$_{2}$As$_{2}$. $T_{1}$ was obtained from fitting $M_z(t) = M_0 \left[ 1-f(0.9 e^{-6t/T_1} + 0.1 e^{-t/T_1} ) \right]$ to the recovery of the nuclear magnetization after inversion \cite{stretchT1}.

\begin{figure}[tbp]
	\centering
	\includegraphics[width=\columnwidth]{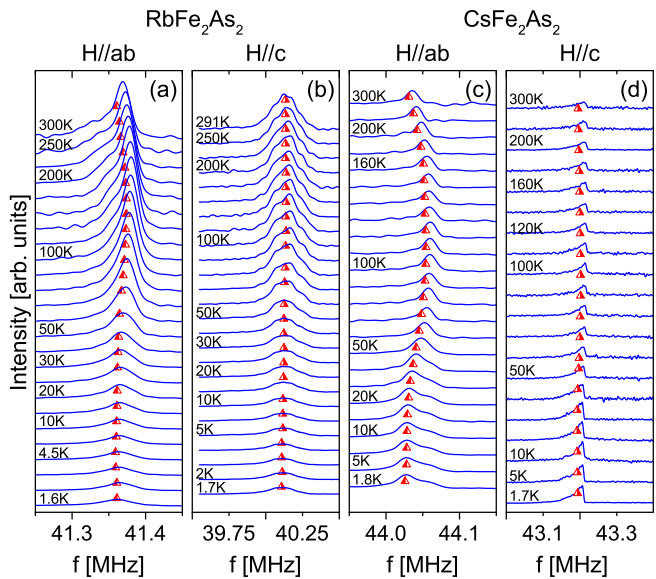}
	\caption{\label{Fig_Spectra}(a) - (d): The $^{75}$As central transition spectra for RbFe$_{2}$As$_{2}$ and CsFe$_{2}$As$_{2}$ with the magnetic field applied in-plane ($H\parallel{ab}$) and out-of-plane ($H\parallel{c}$). Red triangles mark the first spectral moment.}
\end{figure}

By measuring the central ($I_z = -1/2 \rightarrow +1/2$ ) and 
satellite ($-3/2 \rightarrow -1/2$, $+1/2 \rightarrow +3/2$) transition frequencies for $H\parallel{ab}$ and $H\parallel{c}$, we determined the principal axis of the local electric field gradient (EFG) tensor $V_{ZZ}$ to be parallel to the $c$ axis, and the quadrupole frequency $\nu_{Q} \propto V_{ZZ}$  at 5 K as 14.4 MHz for RbFe$_{2}$As$_{2}$ and 13.7 MHz for CsFe$_{2}$As$_{2}$, larger than $\nu_{Q} =$ 12.2 MHz, reported for KFe$_{2}$As$_{2}$ \cite{Wang-K122-pressure}. Between 5 to 300 K, $\nu_{Q}$ shows,
in good accordance with the thermal expansion of the lattice, a weak variation ($\lesssim1\%$) for both RbFe$_{2}$As$_{2}$ and CsFe$_{2}$As$_{2}$, thus proving the absence of a structural transition \cite{K122-crossover,Meingast2012}.
The $^{75}$As central-line spectra of RbFe$_{2}$As$_{2}$ and CsFe$_{2}$As$_{2}$ for $H\parallel{ab}$ and $H\parallel{c}$ are shown in Fig. \ref{Fig_Spectra}. The full width at half maximum (FWHM) is, except for RbFe$_{2}$As$_{2}$ with $H\parallel{c}$, $\approx 30$ kHz and only weakly temperature dependent \cite{Misorientation-FWHM-broadening}. This is even considerably lower than the FWHM of $50-80$ kHz reported for KFe$_{2}$As$_{2}$ \cite{Hirano-jpsj,Wang-K122-pressure}, proving the very high quality of the single crystals used for our NMR study \cite{TdepIntRb122}.

\begin{figure}[tbp]
	\centering
	\includegraphics[width=\columnwidth]{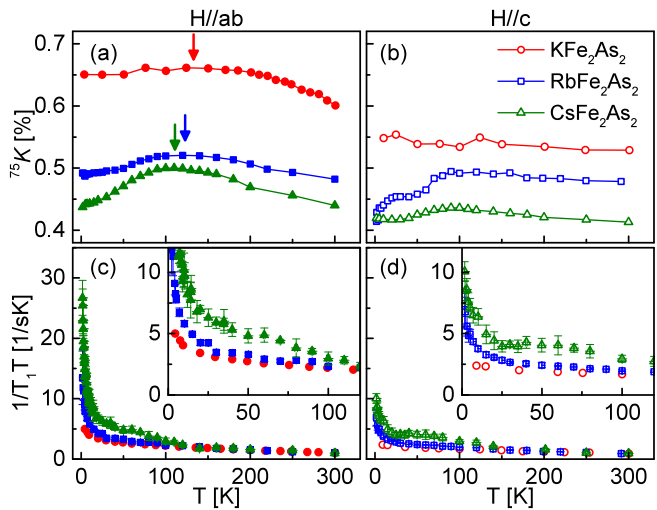}
	\caption{\label{Fig_Kspin} NMR data of RbFe$_{2}$As$_{2}$ and CsFe$_{2}$As$_{2}$ reported in this work, data for KFe$_{2}$As$_{2}$ are from Ref. \onlinecite{Hirano-jpsj}. (a) and (b): Temperature dependence of the Knight shift for $H\parallel{ab}$ and $H\parallel{c}$, the arrows indicate the respective maxima at $T^*$. (c) and (d): Temperature dependence of $1/T_{1}T$ for $H\parallel{ab}$ and $H\parallel{c}$, the insets show zooms of the low-temperature data.}
\end{figure}

The Knight shift $K$ probes the uniform local susceptibility. It is defined by 
$f_{res} = (1 + K) f_{0}$, where $f_{0}={\gamma_{n}}\mu_0H/2\pi$ is the Larmor frequency of a bare nucleus with gyromagnetic ratio ${\gamma_{N}}$ in a magnetic field $H$, and $f_{res}$ is the observed resonance frequency, which we 
extract as the first moment from the recorded spectra. For $H\parallel{ab}$ ($\perp{V_{ZZ}}$), a second-order quadrupole shift $3\nu_{Q}^{2}/16f_{0}$ contributes to $f_{res}$, and was subtracted from the data using the quadrupole frequencies specified above.
As shown in Fig. \ref{Fig_Kspin} (a) and (b), we find a similar anisotropy for all three $A$Fe$_{2}$As$_{2}$ compounds. In general, the in-plane Knight shift $K_{ab}$ is slightly larger than the out-of-plane shift $K_{c}$, and both continuously decrease from KFe$_{2}$As$_{2}$ to CsFe$_{2}$As$_{2}$. This trend indicates a reduction of the uniform local spin susceptibility, in compatibility with an increase of antiferromagnetic correlations. Also, $K_{ab}$ shows a stronger temperature dependence than $K_{c}$, yielding a broad maximum that shifts to lower temperatures and becomes increasingly pronounced.  

In line with the 
Knight shift data, measurements of the macroscopic susceptibility of KFe$_{2}$As$_{2}$ revealed a
maximum around 100 K \cite{K122-crossover}, and are considered as evidence for a coherence-incoherence crossover mechanism \cite{K122-crossover,Liu-Lifshitz-crossover, Wu-XHChen}. It is suggested that localized spins dominate the susceptibility in the incoherent state at high temperatures, whereas the low-temperature coherent state is a metal with enhanced Pauli susceptibility.
By defining $T^{*}$ as the 
maximum of the temperature-dependent Knight shift for $H\parallel{ab}$, we obtain $T^{*} = $133$\pm$3 K, 123$\pm$2 K, and 111$\pm$2 K
for KFe$_{2}$As$_{2}$, RbFe$_{2}$As$_{2}$, and CsFe$_{2}$As$_{2}$, respectively.

Turning to the dynamic correlations, 
$1/T_{1}T$ probes the fluctuating hyperfine fields at the nuclear site, $1/T_{1}T \propto \sum_{q, \alpha, \beta} F_{\alpha \beta}(q) \chi_{\alpha \beta}^{\prime\prime} (q,f_{res})/f_{res}$, with $\alpha, \beta = \left\{ x, y, z \right\}$.  Here, $F_{\alpha \beta}$ denotes the hyperfine form factors and $\chi_{\alpha \beta}^{\prime\prime}$ is the imaginary part of the dynamical electronic susceptibility. The low-temperature part of $(1/T_{1}T)$  increases drastically along the $A$Fe$_{2}$As$_{2}$ series, as shown in Fig. \ref{Fig_Kspin}(c) and (d). This reveals a significant increase of both in-plane and out-of-plane hyperfine-field fluctuations and, therefore, gives strong indications of an increasing dynamical susceptibility upon approaching a magnetic
instability.
Note that for all $A$Fe$_{2}$As$_{2}$ compounds, $(1/T_{1}T)_{H\parallel{c}}$ is generally smaller than $(1/T_{1}T)_{H\parallel{ab}}$.

As was reported for BaFe$_2$As$_2$
and related compounds with K, Co, or Cu doping, the anisotropy ratio of the nuclear spin-lattice relaxation rate, $R = (1/T_{1})_{H\parallel{ab}} / (1/T_{1})_{H\parallel{c}}$,
allows to conclude on the $q$ modulation of electronic spin fluctuations \cite{Hirano-jpsj, Grafe2014}. As detailed in Refs. \onlinecite{Kitagawa2010,Hirano-jpsj},
assuming that the hyperfine field at the As site is given as the sum 
of the fields from the nearest-neighbor Fe spins with tetragonal symmetry (in-plane isotropy of diagonal hyperfine coupling terms, i.e., $A_{ab} = A_a = A_b$), the anisotropy ratio $R$ is given as
\begin{equation}
R=\left\{\begin{array}{lll} 0.5 + 0.5 \left( \dfrac{A_cS_c}{A_{ab}S_{ab}}\right)^2 &\text{no correlation}& \\ 0.5 &(\pi,\pi)&\\
0.5 + \left( \dfrac{S_{ab}}{S_{c}}\right)^2 &(0,\pi) \hspace{1mm} \text{or} \hspace{1mm} (\pi,0).&
\end{array}\right.
\end{equation}
Here, $S_{ab}$ and $S_c$ are the in-plane and out-of-plane
components of the fluctuating Fe spins. For our analysis of $R$, we consistently use the ratio $A_c / A_{ab} = 0.5$, reported for KFe$_2$As$_2$ \cite{Hirano-jpsj}. Since all three members of the $A$Fe$_{2}$As$_{2}$ series are isovalent and the structural parameters vary only slightly \cite{QCP-Eilers}, the hyperfine coupling and chemical shift can be taken as invariant. By further assuming $S_{ab} = S_c$, $R \approx 0.6$ indicates an absence of correlations, $R = 0.5$ corresponds to checkerboard-type fluctuations with $q = (\pi,\pi)$, and $R = 1.5$ results from a fully developed commensurate stripe-type [$q = (0,\pi)$ or $q = (\pi,0)$] modulation.

\begin{figure}[tbp]
	\centering
	\includegraphics[width=\columnwidth]{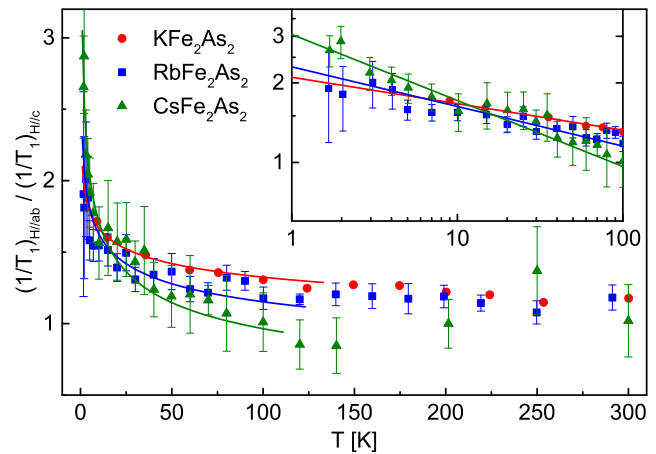}
	\caption{\label{Fig_Korringa}  Temperature-dependent anisotropy ratio $R = (1/T_{1})_{H\parallel{ab}} / (1/T_{1})_{H\parallel{c}}$ for all three $A$Fe$_{2}$As$_{2}$ compounds, data for KFe$_{2}$As$_{2}$ are from Ref. \onlinecite{Hirano-jpsj}. The trend of $R$ is indicated by solid lines as a guide to the eye. The inset shows a zoom of the low-temperature data in log-log scale.}
\end{figure}

Lee et al. reported two-dimensional incommensurate spin fluctuations in KFe$_{2}$As$_{2}$ at $\left(\pi (1 \pm 2 \delta), 0 \right) $, with $\delta = 0.16$, determined by neutron scattering \cite{Lee2011}.
For all three $A$Fe$_{2}$As$_{2}$ compounds, we find, within experimental error, $R \approx 1.2$ at 300 K, see Fig. \ref{Fig_Korringa}. The same room temperature value of the anisotropy ratio was reported for Cu- and Co-doped BaFe$_{2}$As$_2$ \cite{Grafe2014}. Therefore, we conclude that stripe-type fluctuations, in full compatibility with the aforementioned neutron-scattering results are already well developed at room temperature. Further, the presence of these fluctuations implies that the proximite ordered state is not a purely magnetic Mott phase, as this would yield a checkerboard-type modulation. As a general trend, $R$ increases towards low temperatures. For KFe$_{2}$As$_{2}$ and RbFe$_{2}$As$_{2}$,
no change of $R$ is found at $T^{*}$, even though the temperature-dependent exponent of $(1/T_{1})$ for both applied field directions changes at $T^{*}$,
as will be discussed below.
For CsFe$_{2}$As$_{2}$, the $1/T_{1}T$ data
in Figs. 2 (c) and (d) show a hump-like structure at temperatures below about 100 K, which also seems to be reflected by a weak variation of $R$ at intermediate temperatures.
Towards lowest temperatures, $R$ increases to $\approx$ 2 for RbFe$_{2}$As$_{2}$, and to $\approx$ 3 for CsFe$_{2}$As$_{2}$.

As was shown for the case of Cu- and Co-doped BaFe$_{2}$As$_2$, in the vicinity to a quantum critical point, possibly coupled to a structural instability, $R$ tends to increase to much larger values than 1.5 \cite{Grafe2014}. In the present case of the $A$Fe$_{2}$As$_{2}$ compounds, we have no experimental indication of a nearby structural transition. Therefore, an increase of $R$ beyond 1.5 can only result from an increase of $S_{ab}/S_{c}$, proving the increasingly
anisotropic character of correlations close
to the QCP.
Moreover, these findings are compatible with an electronic localization in specific Fe $3d$ orbitals. LDA + dynamical mean field theory (DMFT) calculations have predicted an increasing localization in Fe $3d_{z^{2}}$ and $3d_{xy}$ orbitals in the $A$Fe$_{2}$As$_{2}$ series \cite{Backes-KRbCs-Hund}. In line with these results, the study of Eilers et al. attributes a strong enhancement of effective quasiparticle masses to the hybridization of the $3d_{xy}$ orbitals \cite{QCP-Eilers}.

An increase of anisotropic dynamic correlations leads to an anisotropic power-law behavior with $1/T_{1} \propto T^{\eta}$ close to quantum criticality. Considering the given stripe-type modulation, in-plane spin fluctuations $S_{ab}$ contribute to $(1/T_{1})_{H\parallel{ab}}$ via off-diagonal hyperfine coupling terms, but not to $(1/T_{1})_{H\parallel{c}}$  \cite{Kitagawa2010,Smerald2011,Dioguardi2016}. In consequence, mostly the power-law exponent $\eta_{H\parallel{ab}}$ should evolve upon approaching a QCP driven by in-plane correlations.

\begin{figure}[tbp]
\centering
\includegraphics[width=\columnwidth]{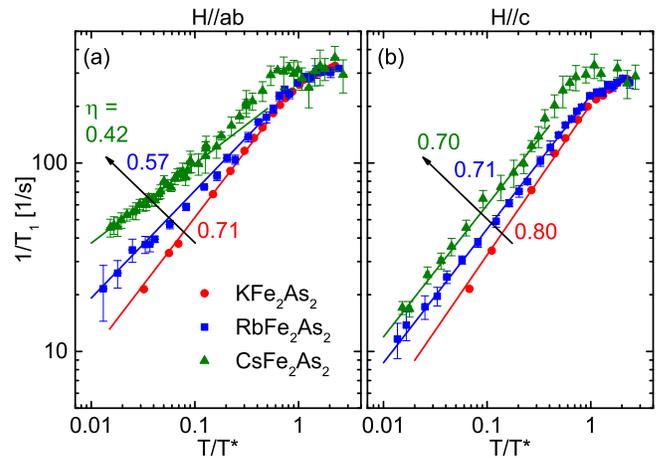}
\caption{\label{Fig_1overT1_Tstar} The spin-lattice relaxation rate $1/T_{1}$ as a function of the reduced temperature $T/T^{*}$ for $A$Fe$_{2}$As$_{2}$ ($A$ = K, Rb and Cs) for (a) $H\parallel{ab}$ and (b) $H\parallel{c}$. The straight lines are fits with $\propto T^{\eta}$ to the low-temperature part. Data for KFe$_{2}$As$_{2}$ are from Ref. \onlinecite{Hirano-jpsj}.}
\end{figure}

Along this line, we now discuss the emergent non-Fermi-liquid character of
the dynamic magnetic correlations by comparing $1/T_{1}$ for the $A$Fe$_{2}$As$_{2}$ compounds for both field orientations. For better comparability, Fig. \ref{Fig_1overT1_Tstar} shows $1/T_{1}$ as a function of the reduced temperature $T/T^{*}$  \cite{Power-T*}. As shown in Fig. \ref{Fig_1overT1_Tstar}(b), all $(1/T_{1})_{H\parallel{c}}$ curves follow almost the same power-law relation with $\eta_{H\parallel{c}}$ between 0.80 and 0.70  below $T^{*}$. For in-plane fields, however, $\eta_{H\parallel{ab}}$ successively decreases from 0.71 to 0.42 below $\approx 0.2 ~ T/T^{*}$.

In Ba$_{0.3}$K$_{0.7}$Fe$_{2}$As$_{2}$, discussed in Ref. \onlinecite{WQYu-K122}, $\eta$ is still close to unity and isotropic, as expected for a Fermi liquid. For KFe$_{2}$As$_{2}$, the reported values of $\eta$ range from 0.75 to 0.8, indicating enhanced spin fluctuations in contrast to a standard Fermi-liquid behavior \cite{WQYu-K122,K-power0.8,Wu-XHChen}. In the present work, we find that $\eta$ becomes increasingly anisotropic for RbFe$_{2}$As$_{2}$ and CsFe$_{2}$As$_{2}$. Moreover, the decrease of $\eta_{H\parallel{ab}}$ evidences that the progressive deviation from Fermi-liquid behavior is mainly reflected by the evolution of the in-plane spin fluctuations $S_{ab}$. 
Our findings for $\eta_{H\parallel{c}}$ are, therefore, 
in general agreement with the results reported by Wu et al., 
but the strong evolution of $\eta_{H\parallel{ab}}$ is
in sharp contrast to their statement on universality 
below $T^*$ \cite{Wu-XHChen}.

In summary, we report a detailed investigation of the anisotropic magnetic correlations in the $A$Fe$_{2}$As$_{2}$ ($A$ = K, Rb, Cs) series by means of $^{75}$As NMR spectroscopy. The Knight-shift data indicate a successive reduction and increasing temperature dependence of the uniform spin susceptibility. 
The anisotropy of $1/T_{1}$ allows to conclude on a stripe-type modulation of spin fluctuations up to room temperature, implying that the proximite ordered state is not a purely magnetic Mott phase. Further, we find an increasing low-temperature anisotropy of spin fluctuations with increase of the alkali-ion radius. The power-law behavior of $(1/T_{1})_{H\parallel{ab}}$ gives evidence that the dynamic correlations evolve towards an increasing deviation from Fermi-liquid behavior, reflected mostly by the in-plane spin fluctuations. All these findings clearly indicate that the electronic correlations in the $A$Fe$_{2}$As$_{2}$ series evolve towards a quantum critical point at the transition to a yet unobserved, ordered phase.
The associated unusual phenomenology, i.e. a decrease of $T_c$ in proximity of the QCP as well as indications for multicomponent order, call for further investigations, in particular, finding an experimental avenue to stabilize the ordered phase.


We thank S.-L. Drechsler, B. B\"{u}chner, and S. Aswartham for valuable
discussions. 
Z.T.Z. was financially supported by the National Nature Science Foundation of China (Grant No. 11304321) and by the International Postdoctoral Exchange Fellowship Program 2013 (Grant No. 20130025). Further, support by the HLD at 
HZDR, a member of the European Magnetic Field Laboratory, and by the Deutsche Forschungsgemeinschaft (DFG) through the Priority Programme SFB 1458 and the Research Training Group GRK 1621 is gratefully acknowledged. 


\begin{thebibliography}{99}
\bibitem{stretchT1} For temperatures below 4 K, a stretching exponent was used for the T$_1$ fitting of RbFe$_{2}$As$_{2}$.
\bibitem{Misorientation-FWHM-broadening}
 The FWHM of RbFe$_{2}$As$_{2}$ for $H\parallel{c}$ is $\approx 200$ kHz due to a residual misorientation of 
 $\approx 10 ^{\circ}$. The related second order quadrupole shift of -6400 ppm was subtracted from the Knight shift data.
\bibitem{TdepIntRb122}
For RbFe$_{2}$As$_{2}$, a loss of the normalized intensity below $\approx$ 100 K indicates reduced sample penetration by the rf field and is attributed to an increase of the surface conductivity at low temperatures \cite{Zhang-RbNonfermi}.
\bibitem{Power-T*}
Note that the power-law exponent of $(1/T_{1})\propto T^{\eta}$ is independent of the choice of $T^{*}$.
\end{thebibliography}
\end{document}